# Electro-optically modulated polarization mode conversion in lithium niobate ridge waveguides

Guang Yang, Alexander V. Sergienko, and Abdoulaye Ndao*

*Department of Electrical and Computer Engineering & Photonics Center, Boston University, 8 Saint Mary's Street, Boston, Massachusetts 02215, USA*

\* andao@bu.edu

**Abstract**

Lithium niobate on insulator (LNOI) waveguides, as an emerging technology, have proven to offer a promising platform for integrated optics, due to their strong optical confinement comparable to silicon on insulator (SOI) waveguides, while possessing the versatile properties of lithium niobate, such as high electro-optic coefficients. In this paper, we show that mode hybridization, a phenomenon widely found in vertically asymmetric waveguides, can be efficiently modulated in an LNOI ridge waveguide by electro-optic effect, leading to a polarization mode converter with 97% efficiency. Moreover, the proposed device does not require tapering or periodic poling, thereby greatly simplifying the fabrication process. It can also be actively switched by external fields. Such a platform facilitates technological progress of photonics circuits and sensors.

## 1. Introduction

The pursuit of low-loss waveguides and on-chip optical components has been at the heart of the development of photonic integrated circuits. Lithium niobate offers an ideal platform due to its broad optical transparency, versatile nonlinear coefficients, and excellent electro-optic properties. Lithium niobate waveguides have been developed using conventional technologies such as proton-exchange [1] and titanium diffusion [2, 3]. However, such embedded waveguides have low index contrast and weak mode confinement, which hinders on-chip applications. Recently, the breakthrough in nanofabrication technologies has made low-loss waveguides possible in lithium niobate on insulator (LNOI) [4-6]. This new platform, featuring high index contrast and tight optical confinement [7], allows small footprints for on-chip components, as well as enables tightly bent structures [8, 9] and ring resonators [6, 10, 11]. Moreover, the compactness of LNOI waveguides greatly facilitates electro-optic modulation, since electrode pairs can be placed much closer, enhancing the efficiency of modulation. A variety of electro-optic applications in LNOI platforms have been demonstrated [12-14], among which a polarization converter, as a highly needed building block for photonic circuits, has been illustrated in [15]. This polarization converter is based on quasi-phase matching between two polarizations. While delivering great performance, this converter requires periodic poling. This increases the fabrication challenge, since periodic poling is typically done prior to waveguide fabrication to select the region with best poling quality [16].

Polarization converter designs in LNOI waveguides have also been proposed in a taper [17]. This design utilizes mode hybridization, a phenomenon widely found in waveguide geometries where the symmetry along one axis, primarily the vertical axis is broken [18]. Tapered polarization converters based on hybrid modes have also been reported previously in other integrated platforms such as indium phosphide [19] and silicon on insulator [20]. Despite proposed methods, a device combining mode conversion and dynamic polarization modulation using ridge waveguides without tapering remains elusive. Here, we propose a novel mode conversion and polarization modulating scheme in LNOI ridge waveguides by introducing an electro-optically perturbed index profile. We show the dynamic modulation of hybrid modes by external electric field and numerically demonstrate a polarization mode converter with 97% efficiency. This paves a new way for sophisticated photonic circuits and sensors that exploit the polarization degree of freedom [21, 22].

## 2. Waveguide design

The proposed device is illustrated in Fig. 1. The ridge waveguide is made in a 0.7 µm thick X-cut lithium niobate thin film with an etching depth of 0.6 µm. The lithium niobate thin film is bonded on a silicon dioxide layer to form the index contrast needed for optical guiding. The top cladding is air. The ordinary and extraordinary refractive indices of lithium niobate at the designed wavelength of 775 nm and room temperature (20 ℃) are calculated to be 2.2586 and 2.1786, respectively, according to the Sellmeier equations in [23]. Mode conversion inside the waveguide is modulated by a transverse electric field in the x direction (see coordinates indicated in Fig. 1). The electric field is assumed to have a longitudinal gradient, leading to an index perturbation profile which also varies longitudinally, as will be elucidated later in the paper.

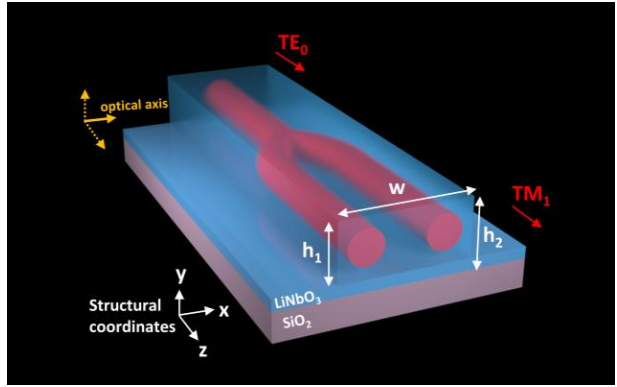

**Figure 1.** A schematic of the proposed device. The lithium niobate layer thickness is $h_2$=0.7 µm. The considered etching depth is $h_1$=0.6 µm. The width $w$ is a parameter under design and is chosen to be 1.1 µm in the final device. The optical axis of the X-cut lithium niobate and the structural coordinates are indicated.

To design such a device, we first solve the waveguide in a 2D cross section without perturbing the refractive index, and perform parameter sweeping to identify a configuration leading to mode hybridization. We used a fully vectorial finite difference mode solver (Lumerical MODE) for our simulation. The effective indices of the first four modes are depicted in Fig. 2(a) in a waveguide with a fixed etch depth of 0.6 µm, while the waveguide width is varied. We found two hybrid modes (hereafter referred to as $Hyb_1$ and $Hyb_2$) between $TE_0$ and $TM_1$ around the waveguide width of $w$=1.09 µm, and the mode profiles of $Hyb_1$ and $Hyb_2$ are visualized in Fig. 2(b)-(d) and (e)-(g), respectively. The intensity profiles of these two modes are almost identical, and their corresponding x(horizontal)- and y(vertical)- components of the electric field display the profile of $TE_0$ and $TM_1$, respectively. To evaluate the degree of hybridization, we define the TE polarization fraction as follows:

$$\text{TE polarization fraction} = \frac{\int |E_x|^2 \mathrm{d}x\mathrm{d}y}{\int \left(|E_x|^2 + |E_y|^2\right) \mathrm{d}x\mathrm{d}y}. \tag{1}$$

This parameter characterizes the fraction of transverse field power that is TE polarized. The corresponding TE polarization fractions of the modes Hyb$_1$ and Hyb$_2$ in Fig. 2(b) and (d) are calculated to be both around 50%, i.e., the two modes are maximally hybridized between TE and TM at $w$=1.09 µm, with their powers composed of equal parts of TE and TM components. Given the almost identical field and polarization composition of the two hybrid modes, their effective indices closely coincide while avoid crossing, as manifested in the inset of Fig. 2(a). This anti-crossing behavior distinguishes hybrid modes from an accidental intersection of two orthogonally polarized modes, in which case the two effective index curves can directly cross each other and form a genuine degenerate point [24]. Therefore, these anti-crossings can be used to identify hybrid regions in parameter sweeps. It is worth noting that mode hybridizations only occur between an even-order TE mode and an odd-order TM mode, or vice versa, if the hybridization is induced by broken vertical symmetry, as is the case for ridge waveguides [17]. To illustrate the mode converter device, the proposed geometrical parameters are $w$=1.1 µm and $h_1$= 0.6 µm.

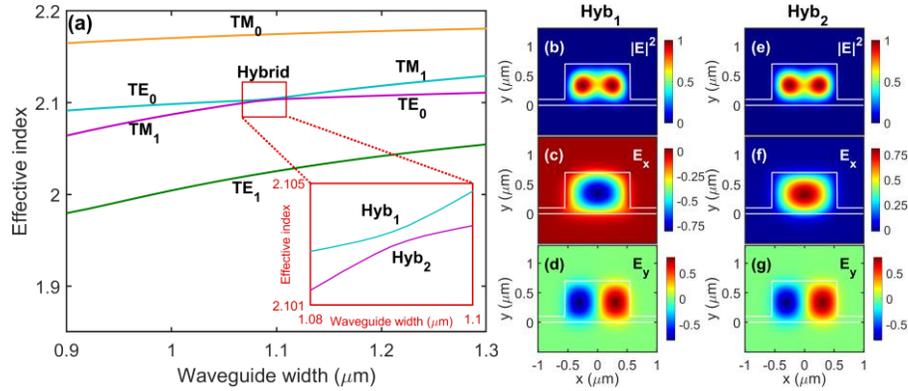

**Figure 2.** Waveguide modes without index perturbation. (**a**) The effective index as a function of the waveguide width for the considered modes at the wavelength of 775 nm. Inset: enlarged view of the hybrid region. (**b**)-(**d**): Mode profiles of Hyb$_1$. (**e**)-(**g**): Mode profiles of Hyb$_2$. (**b**) and (**e**) show the intensity distribution; (**c**), (**d**), (**f**) and (**g**) represent the electric field.

## 3. Performance simulation

We demonstrate a polarization mode converter by modulating the mode hybridization behavior using external electro-optic perturbation of the index profile. We show that with the strong electro-optic effect in lithium niobate, the wavelength at which mode hybridization occurs can be actively controlled. This allows for mode conversion without tapering or quasi-phase matching.

**3.1 Modulation of mode hybridization**

First, we analyze the variation of the mode hybridization behavior when the refractive index is perturbed by a modulating electric field coinciding with the optical axis of lithium niobate (see coordinates in Fig. 1). This configuration engages $r_{33}$, the largest electro-optic coefficient in lithium niobate. The external field $E_{x(\text{ext})}$ would induce a change in the extraordinary refractive index,

$$\Delta n_e = -\frac{n_e^3}{2} r_{33} E_{x(\text{ext})}. \tag{2}$$

However, with the presence of another non-zero component $r_{13}$ in the electro-optic coefficient tensor of lithium niobate, the ordinary index is also perturbed by

$$\Delta n_o = -\frac{n_o^3}{2} r_{13} E_{x(\text{ext})}. \tag{3}$$

We used the experimental values for $r_{33}$ and $r_{13}$ reported in [25]. The change in the ordinary index is noticeably weaker than that in the ordinary index by a constant ratio of $\Delta n_o=0.3722\Delta n_e$ at the designed wavelength. When both indices are perturbed simultaneously, the modes switch between $TE_0$ and $TM_1$ via the two hybrid modes they form, namely $Hyb_1$ and $Hyb_2$ (Fig. 3(a)). For the considered mode conversion, the initial state (state A) is $TE_0$, as indicated by ~97% TE polarization fraction (see Fig. 3(a)), and by the intensity distribution in Fig. 3(b). Wavelength sweeping under the initial condition ($\Delta n_o=\Delta n_e=0$) shows that this state is on the edge of the hybrid region, as manifested by Fig. 3(c). The hybrid region spans a spectral range of ~10 nm, within which the polarization state varies rapidly. This confirms high sensitivity to perturbations. As the refractive indices are perturbed, the hybrid wavelength shifts. Consequently, the initial $TE_0$ mode (state A) at the operation wavelength moves across the hybrid region (state B), and eventually converts to $TM_1$ mode (state C). The relative positions of these states with respect to the hybrid region are represented in Fig. 3 (c), (e) and (g), and the corresponding mode profiles are shown in Fig. 3 (b), (d) and (f). The full dynamic range requires $\Delta n_e=0.0036$ and $\Delta n_o=0.0013$, which correspond to an electric field of 22.2 V/µm. An inverse conversion from $TM_1$ to $TE_0$ (cyan color in Fig. 3(a)) can also be achieved under the same condition.

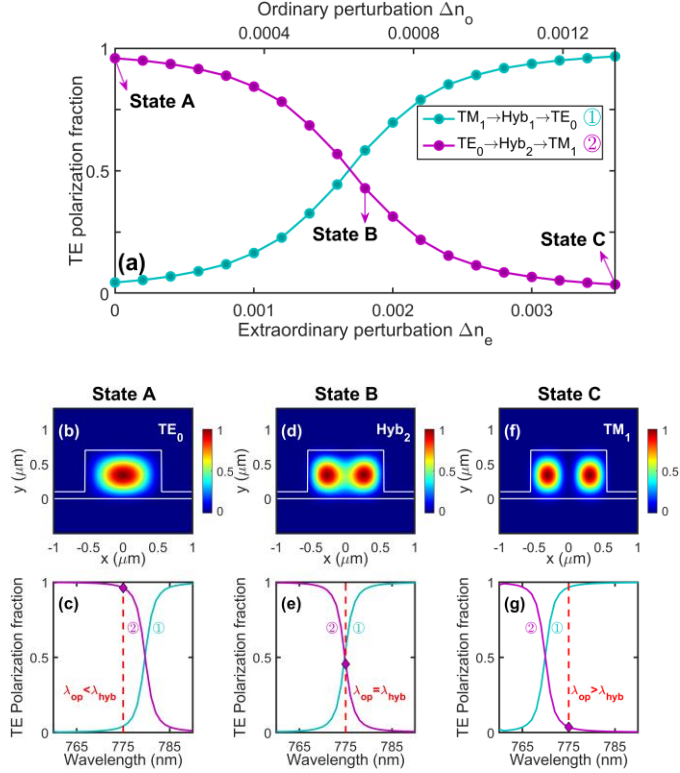

**Figure 3.** Mode switching with varying index perturbation. (**a**) The evolution of modes around hybridization region with varying index perturbation at 775 nm. Bottom and top axes correspond to the perturbation in extraordinary and ordinary indices induced by the same modulating field strength. The intensity profiles of the states A ($TE_0$), B ($Hyb_2$) and C ($TM_1$) indicated in (**a**) are shown in (**b**), (**d**) and (**f**), and the corresponding polarization states with respect to the wavelength are shown in (**c**), (**e**) and (**g**), respectively. $\lambda_{op}$ indicates the operation wavelength (775 nm). $\lambda_{hyb}$ represents the mode hybridization wavelength that shifts from 780 nm (state A) to 770 nm (state C) with varying index perturbation.

## 3.2 Simulation of propagation

It is worth noting that the mode switching described above is considered in a cross section of the waveguide under a static refractive index perturbation. To convert a mode along its propagation, the index perturbation must be implemented longitudinally. Here we demonstrate numerically such a set-up. We consider an electric field profile as shown in Fig. 4(a), and the resultant distributions of extraordinary and ordinary index perturbation are shown in Fig. 4(b) and (c), respectively. In order to realize mode conversion, the perturbed parameter, namely the refractive index here, must be varied slowly. This condition is referred to as "adiabatic", in contrast to a "non-adiabatic" case, where an abrupt change of the perturbed parameter results in insufficient mode conversion [20]. Here we consider an interaction length of 3000 µm, within which the gradient of the electric field, as well as the gradient of the index perturbation, is uniform between z=0 and z=3000 µm. The maximum electric field at z=3000 µm is 22.2 V/µm, corresponding to the field strength required for the final state (state C) in Fig. 3(a). We then inject a $TE_0$ mode in this waveguide and simulate its propagation using the eigenmode expansion (EME) method [26] (Lumerical EME). The EME propagation results for the proposed device are shown in Fig. 4(d)-(f). These profiles are taken from a cross-section in the x-z plane at y=350 µm, corresponding to the center height of the waveguide. The intensity distribution changing from $TE_0$ to $TM_1$ is observed in (d), while (e) and (f), in conjunction, confirm the energy transfer from the x (or TE) polarized component to the y (or TM) polarized component. The conversion efficiency extracted from the scattering matrix is about 96.8%.

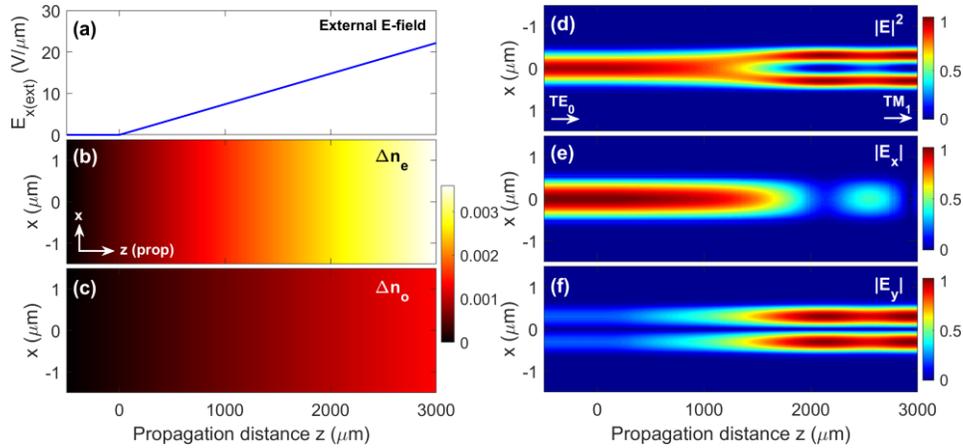

**Figure 4.** Simulated polarization mode conversion when a longitudinally varying electric field is applied. (**a**) The modulating electric field applied in the x direction (or along the optical axis). (**b**) The extraordinary index perturbation profile. (**c**) The ordinary index perturbation profile. (**d**) The intensity profile along propagation. (**e**) The magnitude profile of the horizontal (TE) component $E_x$. (**f**) The magnitude profile of the vertical (TM) component $E_y$. (**d**)-(**f**) correspond to an x-z cross section at y=350 µm.

## 4. Conclusion

In conclusion, we have numerically demonstrated an actively modulated polarization mode converter in an LNOI ridge waveguide. Our proposed scheme is based on refractive index perturbation and avoids the need for changing the waveguide geometry as required by a tapered approach. It also has the advantage of active modulation. Compared to existing methods which are based on quasi-phase matching, our proposed device does not require periodic poling, thereby greatly simplifying the fabrication process. The realization of a such device paves a new way for application in on-chip signal processing and external field sensing.

25. De Toro, J. A., Serrano, M. D., Cabañes, A. G. & Cabrera, J. M. Accurate interferometric measurement of electro-optic coefficients: application to quasi-stoichiometric LiNbO3. *Opt. Commun.* **154,** 23-27 (1998).
26. Sztefka G, & Nolting, H. P. Bidirectional eigenmode propagation for large refractive index steps. *IEEE Photon. Technol. Lett.* **5,** 554-557 (1993).